\documentclass[runningheads]{llncs}
\usepackage{graphicx}
\usepackage{caption}
\usepackage{subcaption}
\usepackage{float}
\usepackage{amsmath}

\usepackage[utf8]{inputenc}
\usepackage[export]{adjustbox}
\usepackage{wrapfig}
\begin{document}
\title{U-Net with Graph Based Smoothing Regularizer for Small Vessel Segmentation on Fundus Image}

\titlerunning{U-Net with Graph Based Smoothing Regularizer}

\author{Lukman Hakim\inst{1} \and
Novanto Yudistira\inst{1} \and
Muthusubash Kavitha\inst{1}\and
Takio Kurita\inst{1}}
\authorrunning{L. Hakim et al.}
%
\institute{Department of Information Engineering, 
Hiroshima University, 1-4-1 Kagamiyama, Higashi-Hiroshima, 739-8527 Japan \\
\email{\{lukman-hakim,kavitha,cbsemaster,tkurita\}@hiroshima-u.ac.jp}}
\maketitle              
\begin{abstract}
The detection of retinal blood vessels, especially the changes of small vessel condition is the most important indicator to identify the vascular network of the human body. Existing techniques focused mainly on shape of the large vessels, which is not appropriate for the disconnected small and isolated vessels. Paying attention to the low contrast small blood vessel in fundus region, first time we proposed to combine graph based smoothing regularizer with the loss function in the U-net framework. The proposed regularizer treated the image as two graphs by calculating the graph laplacians on vessel regions and the background regions on the image. The potential of the proposed graph based smoothing regularizer in reconstructing small vessel is compared over the classical U-net with or without regularizer. Numerical and visual results shows that our developed regularizer proved its effectiveness in segmenting the small vessels and reconnecting the fragmented retinal blood vessels.

\end{abstract}
\section{Introduction}
Characteristics of blood vessels in retina guide an ophthalmologist to diagnose pathologies of different eye anomalies such as age-related macular degeneration (ARMD) and diabetic retinopathy (DR)\cite{jour1}\cite{jour2}.
Additionally, it helps to identify several physiological problems, specially hypertension and some other cardiovascular diseases \cite{jour3}. However, it is time consuming process to identify the disease caused blood vessels, especially the changes of states of small vessel and its characteristics. To rectify the subjective detection of retinal blood vessels, several automated system of segmentation of blood vessels were developed. 
However, the separation of the blood vessel is not an easy task because of the small and fragmented structure in low contrasting retinal image. 

Chakraborti et al \cite{jour4} presented a self- adaptive matched filter by combining the vesselness filter with the matched filter for the detection of blood vessels on the retinal fundus image. 
Tagore et al \cite {jour5} presented a new algorithm for retinal blood vessel segmentation by using the intensity information of red and green channels of color fundus image. It helped to distinguish between vessels and its background in the phase congruency image.
Recently, several researchers have been implementing convolutional neural network (CNN) for retinal blood vessel segmentation. Melinscak \cite{jour6} presented retinal vessel segmentation system using ten layers of CNN.  In \cite{liskowski2016}, a structured prediction scheme was used to highlight the context information, while testing a comprehensive set of architectures. Fu et al \cite{fu2016} combined a typical 7-layer CNN with a conditional random field and reformulated a recurrent neural network to model long-range pixel interactions. Li et al \cite{jour7} considered the vessel segmentation task as a cross-modality data transformation problem in a deep learning model. Dasgupta \cite{jour8} proposed a neural network framework that iteratively classify pixels from the fundus image. 
Although the localization of the vessels has improved  significantly with CNN, the fragmented small vessel identification is still a challenging task. Because sometimes it is located at the end of the vessel branch and failed to maintain the connectivity. Furthermore, it is difficult to detect isolated vessels in the low contrast background.  

For addressing these challenges, first time we proposed a graph based smoothing regularizer that considers the image into two regions by calculating graph laplacians on vessels and its background areas. The proposed regularizer is used as a objective function in the deep CNN framework makes the network can efficiently learn the pixel connectivity of the small or isolated blood vessel structure. The effectiveness of our proposed regularization term was evaluated and compared using U-net architecture and baseline U-net. The performance of the proposed approach was also compared with the state-of-the-art networks model in reconstructing the small and isolated vessel regions.

\begin{figure}[t!]
\includegraphics[width=\textwidth]{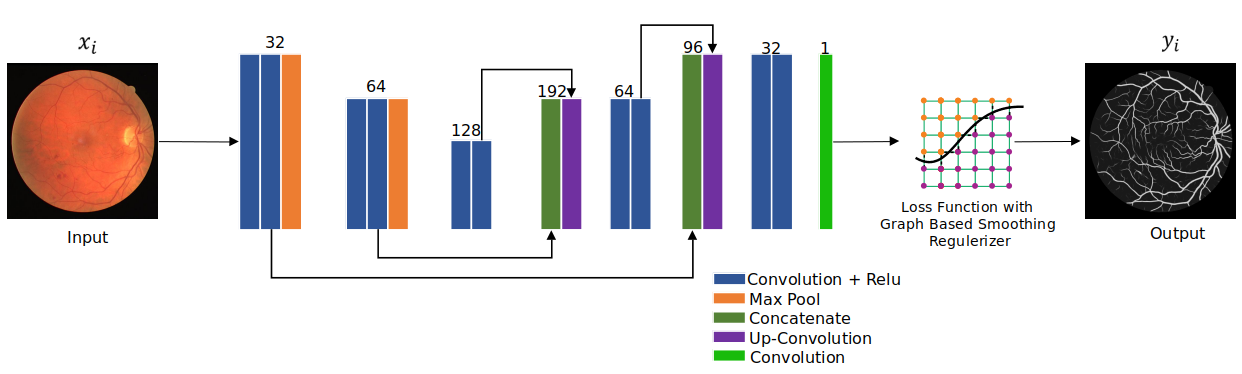}
\caption{U-Net framework using graph based smoothing regularizer  } \label{fig:unet-model}
\end{figure}

\section{Methods}
 
In parallel to track large vessel we are interested to reconstruct the small or isolated vessels. Paying attention to segment the small vessels in the fundus region, we considered to define a regularizer that is based on the graph laplacian smoothing method. We evaluate the effectiveness of our proposed regularizer using U-Net architecture \cite{ronneberger2015}.  The schematic diagram in Fig. \ref{fig:unet-model}, describes the proposed graph based smoothing regularizer for small vessel reconstruction in the U-net framework.

\subsection{Network architecture}
The network has two parts, encoder and decoder module. The encoder module of the network contains 6 convolutional layers with ReLU and 2 max-pooling layer with 32, 64 and 128 feature maps, respectively. The decoder module contains 4 convolutional layers with ReLU and one convolution layer without ReLU. The input feature maps are upsampled with the factor of 2. Skip connections are used to concatenate the feature maps after the deconvolution with the corresponding features from the encoder path. After the decoder, pixel-based probability maps and predictions are generated by a sigmoid classifier function. 
We used data augmentation to boost the training performance of the network. The data augmentation methods using patching were applied on the original data with the patch size of 48x48 pixels. 

\begin{figure}[b!]
\centering
\includegraphics[width=0.8\textwidth]{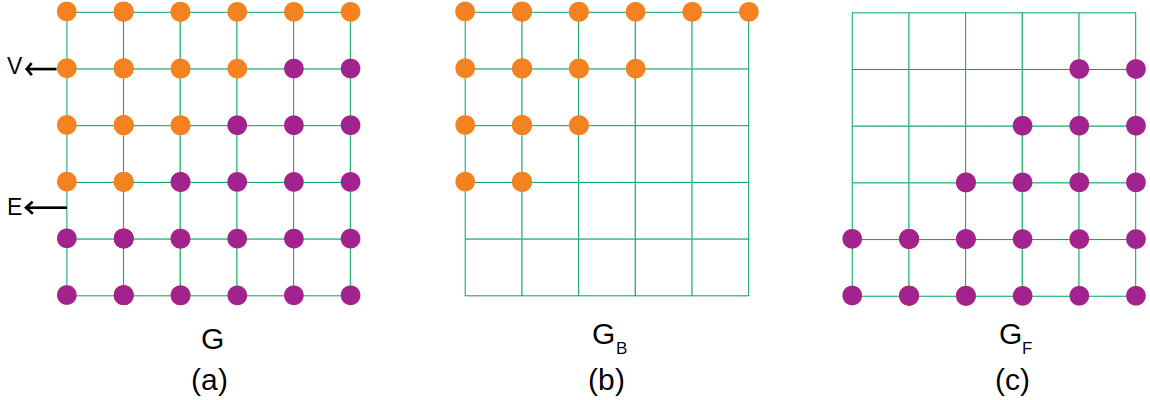}
\caption{Representation of set of nodes of the pixel graph to compute boundary between two regions, a) image as an graph, b) background region graph and c) foreground region graph.}
\label{fig-construct-graph}
\end{figure}

 \subsection{Graph Based Smoothing as Regularizer}
Graph based smoothing regularizer is based on the graph laplacian matrix. Graph laplacian can be obtained by constructing the adjacency graph and diagonal matrix. With the help of graph laplacian, the image pixels can be interpreted as node. Every node  is connected with every other node in the graph. In this study, we formally defined two graphs $G_F$ and $G_B$ for foreground and background, respectively as shown in Fig. \ref{fig-construct-graph}. $G_F$ and $G_B$ includes a pair of $(V_F,E_F)$, $(V_B,E_B)$, respectively. The parameters $V_F$ and $V_B$ are finite set of elements called vertices, and $( E_F=\{(j_F,k_F)|j_F \in V_F, k_F \in V_F\} $ and $(E_B=\{(j_B,k_B)|j_B \in V_B, k_B \in V_B\} $ are edges.   

Let us consider $(x_i,t_i)|i=1,...,M$, where $x_i$ is a $i^{th}$ input data from the training dataset $X$, and $t_i$ is a label from training dataset. The number of training samples and labels is denoted by $M$ and $N$, respectively.

The proposed CNN based U-Net architecture is trained to predict the output image pixels $y_i$ from a given input image pixels $x_i$. For each edge of foreground  $(j_F,k_F) \in E_F$  and background $(j_B,k_B) \in E_B$ of the pixel graph, the similarity  $\beta_{(j_F,k_F)} $ and $\beta_{(j_B,k_B)}$ is defined as 
\begin{equation}
\beta_{(j_F,k_F)} = 1-|t_{j_F}-t_{k_F}|
\end{equation}

\begin{equation}
\beta_{(j_B,k_B)} = 1-|t_{j_B}-t_{k_B}|
\end{equation}

We introduced the regularization term for smoothing $S$ based on foreground region $F$ and background region $B$ as
\begin{equation}
    \sum_{(j_F,k_F) \in G_F}\beta_{j_F,k_F}{(y_{j_F}-y_{k_F})^2}
    = y^T(D_F - A_F)y=y^T L_{F} y
\end{equation}
\begin{equation}
    \sum_{(j_B,k_B) \in G_B}\beta_{j_B,k_B}{(y_{j_B}-y_{k_B})^2}
    = y^T(D_B - A_B)y=y^T L_{B} y
\end{equation}
where, $L_F$ and $L_B$ is Laplacian graph. The adjacency and diagonal matrices is defined as following 
\begin{equation}
    \binom{A_F=\beta_{(j_F,k_F)},   D_F = \sum_{j_F=1}^N \beta_{j_F, k_F}}{A_B=\beta_{(j_B,k_B)},   D_B = \sum_{j_B=1}^N \beta_{j_B, k_B}}
\end{equation}
Smoothing $S$ can be written as 
\begin{equation}
    S = y^T(L_{F}+L_{B}) y \\
    = y^{T}L_{G}y
\end{equation}

The objective function $O$ applied in this study is the summation of the binary cross entropy of each label with the regularization term using graph based smoothing, which is defined as 

\begin{equation}
O = \sum_i^M \{t_{i}log(y_{i})+(1-t_i)log(1-y_i)\}+\lambda \sum_i^M S
\end{equation}

\begin{figure}[b]
\includegraphics[width=\textwidth]{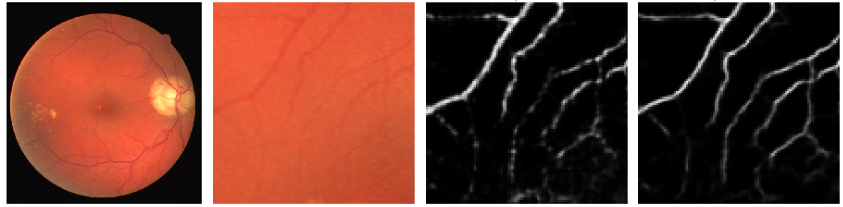}
\caption{Segmenting  small vessels. From left to right: Fundus image, patch image, results of network without regularizer, and with proposed graph based smoothing regularizer } \label{fig6}
\end{figure}

One of the complexity of the graph construction is depends on the size of images. Therefore, we calculate laplacian graph randomly on background and foreground respectively to reduce the complexity.

\section{Experiments}
\subsection{Dataset}
In this study, the proposed method is evaluated on the DRIVE datasets. The DRIVE dataset consists of 40 fundus images. The manual segmentations of the vessels is provided for the all the datasets. The vessels of small width or isolated pixels were defined as small vessels. The size of the image is 565 x 584 pixels with 8 bits per color channel. The number of training and testing data used in this study is 20 and 20, respectively. 
\subsection{Experimental settings}
In order to find a suitable $\lambda$ value for the regularizer, we varied this parameter such as $\lambda=$ 0.0001, 0.00001, 0.000001 and 0.0000001 for accurate small vessel construction. we used Adam optimizer with learning rate of 0.001 and 100 epoch to train the model. we chose binary cross entropy loss because it greatly improved the performance of the model. In our experiments, we compared our proposed network with graph based smoothing regularizer over the baseline U-net without regularizer. The performance of the proposed approach was also compared with the results of the  state-of-the-art networks model. The sensivity (Se), specificity (Sp), accuracy (Acc) and area under the curve (AUC) was used to evaluate the performance of the proposed approach. The  proposed approach is implemented in pytorch with Intel(R) Core(TM) i7-6700K CPU @ 4.00GHz Processor, 32 GB of RAM and Nvidia GeForce GTX 1080/PCIe/SSE2 graphic cards.

\begin{figure}[t!]
\centering
\includegraphics[width=0.8\textwidth]{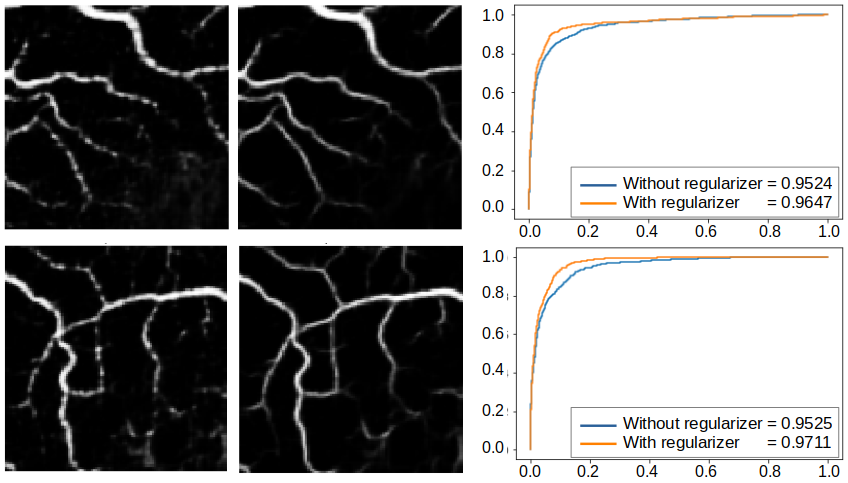}
\caption{Segmented small vessels. From left to right:  without graph based smoothing regularizer, proposed graph based smoothing regularizer and AUC performance} \label{fig-compare-auc}
\end{figure}

\section{Results}
The appropriate regularizer parameter value $\lambda$ to reconstruct the width of the small vessel retained with adequate information was found with the value of 0.000001. The proposed graph based smoothing regularizer coupling with U-net succeeds in reconstructing both large and small vessel pixels compared over U-net without regularizer is presented in Fig. \ref{fig6}. In this figure, the improvement of the disjointed vessel connectivity is clearly observed from the patch based fundus image.

\begin{table}[b!]
\caption{Performance comparison of our approach with state-of-the art methods interns of sensitivity, specificity, accuracy and AUC }\label{tab1}
\begin{center}
\begin{tabular}{|p{2cm}|p{3cm}|p{1cm}|p{1cm}|p{1cm}|p{1cm}|}
\hline
\bf Year & \bf Method & \bf Se & \bf Sp & \bf Acc &  \bf AUC\\
\hline
2016 &  Azzopardi et al. \cite{azzopardi2015} & 0.7655 & 0.974 & 0.9442& 0.9614\\
2016 &  Khan et al. \cite{khan2016}  & 0.7373 & 0.9670 & 0.9501 & -\\
2016 &  Zhao et al. \cite{zao2016} & 0.7420& 0.9820& 0.950& 0.8620\\
2018 &  Marin et al. \cite{marin2018} & 0.7067 & 0.9801& 0.9452& 0.9588\\
2018 &  Orlando et al. \cite{orlando2018} & 0.7897& 0.968 & -&-\\
2016 & Fu et al. \cite{fu2016} & 0.7294 & - & 0.9470 & - \\
2015 & Wang et al. \cite{wang2015} & 0.8173 & 0.9733 & 0.9533 & 0.9475\\
2016 & Liskowski et al. \cite{liskowski2016}& 0.7569 & 0.9816 & 0.9527 & 0.9738 \\
- & U-net \cite{ronneberger2015} &0.6707 & 0.9867&0.9465 &0.9652 \\
- & Proposed method & 0.7064 &\bf 0.9897 & \bf0.9536 & \bf 0.9794 \\
\hline
\end{tabular}
\end{center}
\end{table}

The graph based smoothing regularizer in the U-net depends on the pixel connectivity criterion. Hence, when we compared the AUC value of architecture without regularizer, our approach resulting high AUC value with large number of vessels. The qualitative and quantitative results of our approach in segmenting most of the vessels is shown on the image patch examples (Fig. \ref{fig-compare-auc}).
\begin{figure}[t!]
\includegraphics[width=\textwidth]{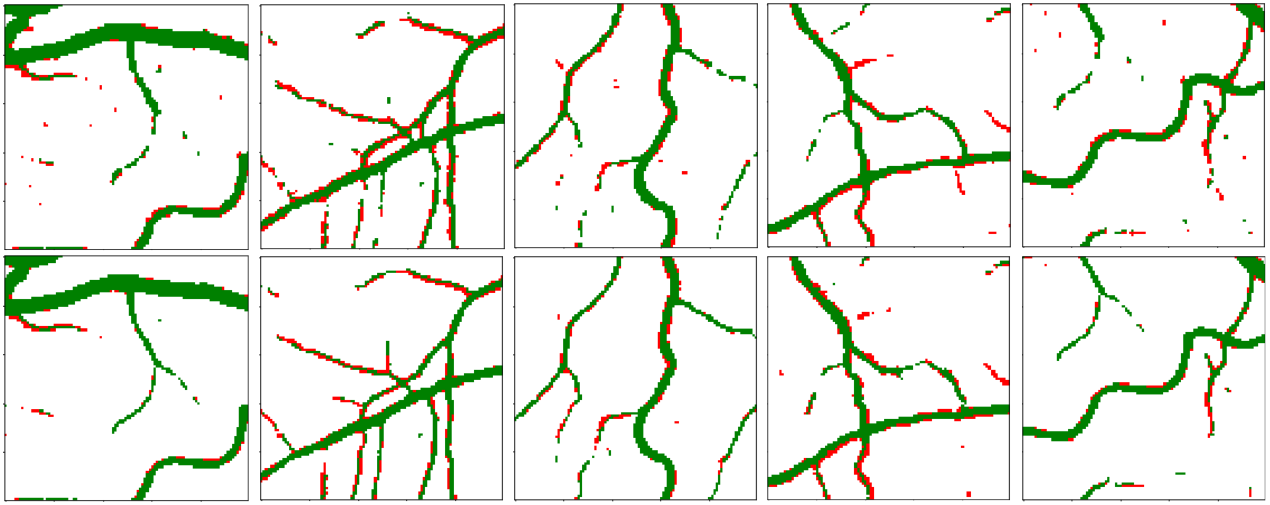}
\caption{Small vessel segmentation examples: Green, and red colors represented TP, and FN respectively. First row shows the segmented vessels without graph based smoothing regularizer and the second row shows the results of the proposed graph based smoothing regularizer.} \label{fig-compare-tp-fn-green-red}
\end{figure}
Our approach achieves significantly higher performance with high AUC value (0.979) in segmenting the small retinal blood vessels than the other  state-of-the-art methods is presented in Table \ref{tab1}    . 
Sensitivity of our method is moderate and it is almost similar with other conventional methods. However, the proposed approach achieves considerably higher specificity (0.99) than all the other methods, which reconstructs to the segmentation of more vessels. When we considered accuracy, our approach scored higher value (0.95), and very close to \cite{wang2015} and \cite{liskowski2016} methods. Fig. \ref{fig-compare-tp-fn-green-red} explains examples of the analysis of vessel reconstruction, where this study focused on the small
vessels. The segmentation results are colorized to demonstrate the confusion matrix: green pixels indicate the TPs and red pixels represent the FNs. Graph based network showed highly acceptable performance for small vessel reconstruction. Our method almost reconnecting all isolated vessels and it can be observed from the lager number of TP pixels. When we analyzed the network without our proposed regularizer, it produced large number of FN pixels. 

\section{Conclusions}
We newly proposed a graph based smoothing regularization term with the loss function in the U-net framework for the segmentation of small vessels in the retinal image. The proposed regularization term effectively computing the  graph laplacians on both vessels and its background regions and thus it significantly reduced the segmentation errors and reconnected small fragmented vessels. Our approach can segment more number of vessels and almost reconnecting all isolated vessels than the baseline U-net without regularizer. Compared to other state-of-the-art methods, our approach demonstarted its improvement in retaining width of the small vessel and disjointed vessel connectivity through its high AUC value. Future work will focus on implementing the proposed regularization term on different retinal image datasets and different segmentation CNN architectures.

\section*{Acknowledgments} 

This work was partly supported by JSPS KAKENHI Grant Number 16K00239 and 18F18112.

\end{document}